\documentclass{article}
\usepackage[dvips]{graphicx}  
\usepackage{color}

\usepackage[varg]{txfonts}
\usepackage{concmath}

\begin{document}

\newcommand{\rum}{\rule{0.5pt}{0pt}}
\newcommand{\rub}{\rule{1pt}{0pt}}
\newcommand{\rim}{\rule{0.3pt}{0pt}}
\newcommand{\numtimes}{\mbox{\raisebox{1.5pt}{${\scriptscriptstyle \rum\times}$}}}
\newcommand{\numtimess}{\mbox{\raisebox{1.0pt}{${\scriptscriptstyle \rum\times}$}}}
\newcommand{\Boldsq}{\vbox{\hrule height 0.7pt
\hbox{\vrule width 0.7pt \phantom{\footnotesize T}%
\vrule width 0.7pt}\hrule height 0.7pt}}
\newcommand{\two}{$\raise.5ex\hbox{$\scriptstyle 1$}\kern-.1em/
\kern-.15em\lower.25ex\hbox{$\scriptstyle 2$}$}

\renewcommand{\refname}{References}
\renewcommand{\tablename}{\small Table}
\renewcommand{\figurename}{\small Fig.}
\renewcommand{\contentsname}{Contents}

\twocolumn[%
\begin{center}
{\Large\bf 
Dynamical 3-Space: Gravitational Wave Detection and the Shnoll Effect\rule{0pt}{13pt}}\par

\bigskip
 David P. Rothall and Reginald T. Cahill\\ 

{\small\it  School of Chemical and Physical  Sciences, Flinders University,
Adelaide 5001, Australia\rule{0pt}{15pt}}\\
\raisebox{+1pt}{\footnotesize E-mail: David.Rothall@flinders.edu.au, Reg.Cahill@flinders.edu.au}\par

\bigskip

{\small\parbox{11cm}{%
Shnoll has  investigated the non-Poisson scatter of rate measurements in various phenomena such as biological and chemical reactions, radioactive decay, photodiode current leakage and germanium semiconductor noise, and attributed the scatter  to cosmophysical factors. While Shnoll didn't pinpoint the nature of the cosmophysical factors the Process Physics model of reality  leads to a description of space, which is dynamic and fractal and exhibits reverberation effects,  and which  offers an explanation for the scattering anomaly. The work presented here shows a new way of generating the effects Shnoll discovered, through studying the phase difference of RF EM waves  travelling through a dual coaxial cable Gravitational Wave Detector experiment.
\rule[0pt]{0pt}{0pt}}}\medskip
\end{center}]{%

\setcounter{section}{0}
\setcounter{equation}{0}
\setcounter{figure}{0}
\setcounter{table}{0}

\markboth{Rothall D.P. and Cahill R.T. Dynamical 3-Space: Gravitational Wave Detection and the Shnoll Effect}{\thepage}
\markright{Rothall D.P. and Cahill R.T. Dynamical 3-Space: Gravitational Wave Detection and the Shnoll Effect}

\section{Introduction - Shnoll Effect}\label{sect:shnoll}
 
Over sixty years ago Simon Shnoll discovered a scatter anomaly in the measurements of the reaction rates of ATP-ase in actomyosin solutions over time that could not be explained \cite{s1}. Extensive research into this scatter anomaly lead to the conclusion that the reaction rates of the protein solution not only varied with time, but followed a distribution with preferred (discrete) values instead of a typical Poisson distribution. Over the following decades it was found that quite different phenomena also displayed similar scatter anomalies, ranging from chemical reactions to $\alpha$-radiation activity in \textsuperscript{239}Pu decay, photomultiplier dark noise and semiconductor noise fluctuations\cite{s2}. Shnoll's investigation of the scatter anomaly (referred to here as the Shnoll effect),  between May 28 - June 01, 2004, produced  352,980 successive measurements of the $\alpha$ decay of a  \textsuperscript{239}Pu source\cite{s1}. Radioactive decay is considered to be a stochastic process, i.e. a random process with no preferred frequencies, and hence follows Poisson statistics. Fig.\ref{fig:shnollplot} is a layer histogram taken from Shnoll's data, with layer lines taken every 6000 measurements.  The y-axis represents the frequency of decay rates and the x-axis is the number of decays per second - the decay rate. Over time the  layer lines of the histogram exhibit  a fine structure which  become more prominent with more measurements, instead of canceling out as in the case of a typical Poisson distribution. This suggests that the radioactivity of \textsuperscript{239}Pu takes on discrete values, and is not completely  random.

\begin{figure}[t]
\centering
\includegraphics[scale=0.55]{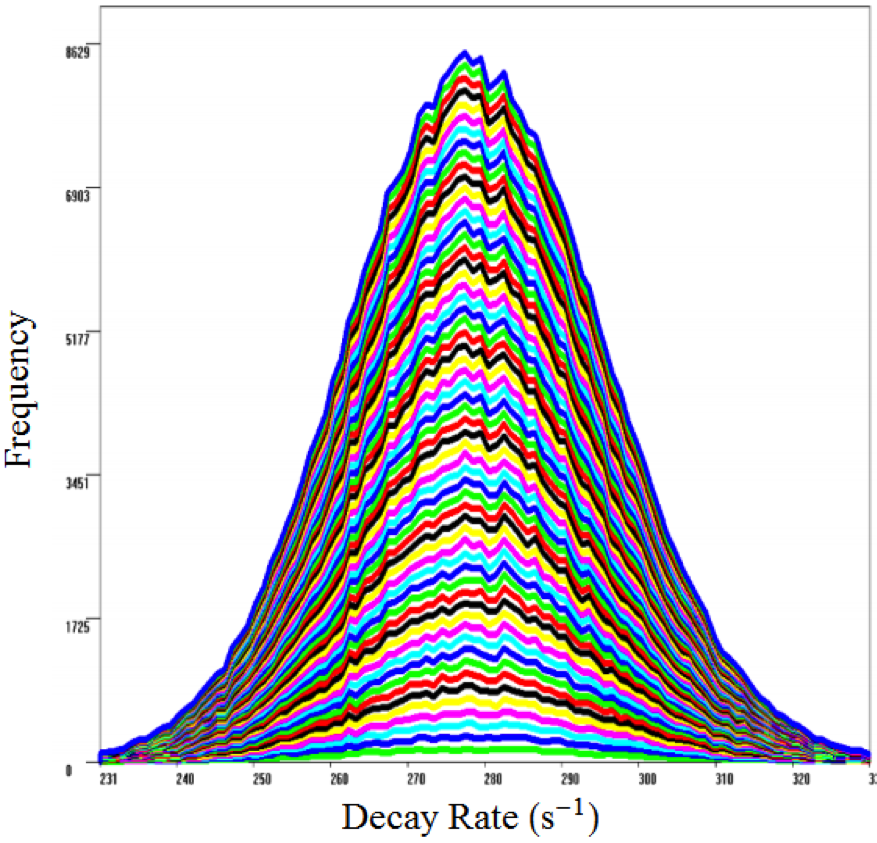}
\vspace{-4mm}\caption{Non-Poisson distribution of 352,980 measurements of \textsuperscript{239}Pu $\alpha$ decay by Shnoll performed in 2004 (Fig. 2-2 of \cite{s1}). The layered histograms are taken every 6000 measurements. The x-axis denotes the number of decay events per second and the y-axis is the frequency of decay rate measurements.}
\label{fig:shnollplot}\end{figure}

Upon further study it was found that not only did the distribution (histogram) shapes vary over time, but the histogram shapes also correlated between different experiments run in parallel, regardless of whether they were located in the same laboratory or separated by thousands of kilometres. This was referred to as absolute time synchronism. Local-time synchronism was also observed, where histogram shapes of one experiment matched those of another with a time delay corresponding to the difference in longitudes of the two locations of the experiments (i.e. as the Earth rotates). A ``near zone" effect was also discovered, where consecutive histograms in time of an individual experiment were found to be most similar in shape, regardless of the time scale used to generate the histograms, indicating the fractal nature of the scattering anomaly. The main conclusions drawn from Shnoll's research was that the consistency of the ``scattering of results" of measurements in a time series arise due to inhomogeneities in the ``space-time continuum" \cite{s1,s3}. These inhomogeneities are ``caused by the movement of an object in the inhomogeneous gravitational field", e.g. as the Earth rotates/orbits the Sun, as the moon orbits the Earth etc. While these inhomogeneities were not characterised by Shnoll there is a remarkable amount of evidence supporting this conclusion.
 
\begin{figure*}[t]
\centering
\includegraphics[scale=0.5]{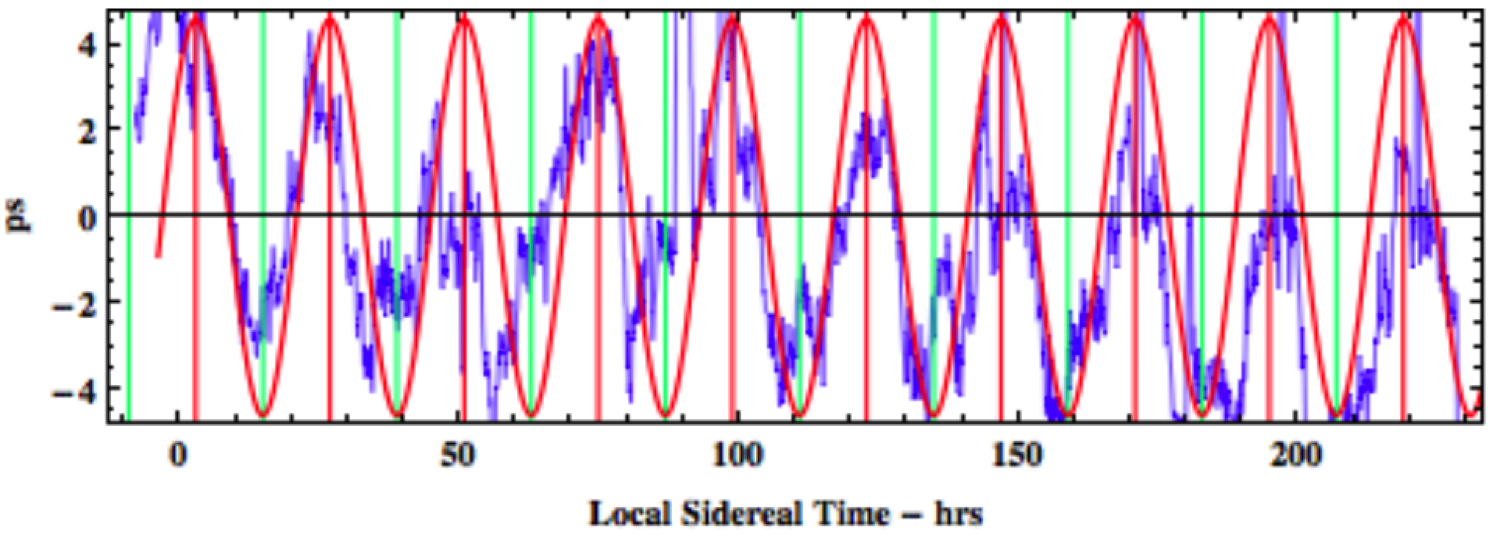}
\vspace{-4mm}\caption{\small{Reproduction of figure 8 (top) from \cite{c1} showing the travel time differences (ps) between the two coaxial cable circuits in \cite{c1} plotted against local sidereal time, for the duration March 4 - 12, 2012. The smooth sine wave is a prediction made from the Dynamical 3-Space theory using NASA spacecraft Earth-flyby Doppler shift data.}
\label{fig:regcoax}}\end{figure*}

\section{Dynamical 3-Space}\label{sect:3space}

An alternative explanation of the Shnoll effect has been proposed  using an alternative theory known as dynamical 3-space theory; see Process Physics  \cite{Book}. This arose from modeling time as a non-geometric process, i.e. keeping space and time as separate phenomena, and leads to a description of space which is itself dynamic and fractal in nature. It uses a uniquely determined generalisation of Newtonian Gravity expressed in terms of a velocity field $\textit{\textbf{v}}(\textit{\textbf{r}},t)$, defined relative to an observer at space label  coordinate  $\textit{\textbf{r}}$, rather than the original gravitational acceleration field. The dynamics of space in the absence of vorticity, $\nabla\times \textit{\textbf{v}}=\textit{\textbf{0}}$, becomes\footnote{The $\alpha$ term in (\ref{eqn:3space}) has recently been changed  due to a numerical error found in the analysis of borehole data. All solutions are also altered by these factors.  (\ref{eqn:3space}) also contains higher order derivative terms - see \cite{UnivBH} .}
\begin{equation}
\nabla\! \cdot\!\left(\frac{\partial \textit{\textbf{v}} }{\partial t}+ (\textit{\textbf{v}}\!\cdot\! \nabla)\textit{\textbf{v}}\right)+
\frac{5 \alpha}{4}\left((tr D)^2 -tr(D^2)\right) =-4\pi G\rho, \nonumber \\
\label{eqn:space}\end{equation}
\begin{equation}
D_{ij}=\frac{\partial v_i}{\partial x_j}
\label{eqn:3space}\end{equation}
where $\rho(\textit{\textbf{r}},t)$ is the usual matter density.
The 1st term involves the Euler constituent acceleration, while the $\alpha-$term describes a significant self interaction of space. Laboratory, geophysical and astronomical data suggest that $\alpha$ is the fine structure constant $\approx 1/137$. This velocity field corresponds to a space flow which has been detected in all of the experiments listed in section \ref{sec:WGen}. In the spherically symmetric case and in the absence of matter $\rho = 0$, (\ref{eqn:3space}) contains solutions for black holes (spatial inflows) and an expanding universe (Hubble expansion) along with that for black holes embedded in an expanding universe \cite{UnivBH}.  Eqn.(\ref{eqn:3space}) also contains solutions for the inflow of space into a matter density. Perturbing the spatial inflow into matter (i.e. simulating gravitational waves) was shown recently to produce reverberations in which the wave generates trailing copies of itself\cite{c2}. This reverberation effect is caused by the non-linear nature of the flow dynamics evident in (\ref{eqn:3space}) and will be shown in the coaxial cable data discussed in section \ref{sec:WGen}.

\section{2012 Dual RF Coaxial Cable Experiment}\label{sec:WGen}
The dynamical 3-space theory was applied to an experiment which studied the radio frequency (RF) electromagnetic (EM) speed anisotropy, or ultimately the absolute motion of Earth through space. The effect of absolute motion has previously been studied using the results from Michelson - Morley, Miller,  and DeWitte experiments \cite{c1}. These results are in remarkable agreement with the velocity of absolute motion of the Earth determined from NASA spacecraft Earth-flyby Doppler shift data all revealing a light/EM speed anisotropy of some 486 km/s in the direction RA=$4.3^{h}$, Dec = $-75.0^{\circ}$ \cite{CahillNASA}. The actual daily average velocity varies with days of the year because of the orbital motion of the Earth - the aberration effect discovered by Miller, but also shows fluctuations over time. The dual RF coaxial cable experiment, performed from March 4 - 12, 2012, measures the travel time difference of two RF signals propagating through dual coaxial cables\cite{c1}. The key effect in this 1st order in $v/c$ experiment is the absence of the Fresnel drag effect in RF coaxial cables at a sufficiently low frequency. The experiment is designed such that one RF signal travels through one type of coaxial cable and returns via another type of cable, while the other signal does exactly the opposite. The cables are bound together such that any travel time effects due to temperature changes cancel as both cables are affected equally. Fig.\ref{fig:regcoax} is a reproduction of the data obtained from the experiment in March 2012 where the travel time difference between the RF signals is plotted against sidereal time. The data is fitted, smooth curve,  using predictions from the NASA  spacecraft Earth-flyby Doppler shift data, where a flow of space traveling at a speed of 499 km/s and direction RA=$2.75^{h}$, Dec = $-77^{\circ}$ predicts the overlaid sine wave, with dynamic range $\sim$8 ps. The Earth rotation effect, wrt the galaxy, can be observed from the data, as well as turbulence effects. Turbulence effects are beginning to be characterised, and can be shown to correspond to what are, conventionally known as gravitational waves, although not those implied by General Relativity, but more precisely are revealing a fractal structure to dynamical 3-space, as illustrated in fig.\ref{fig:space6}.

A fast Fourier transform of the data in fig.\ref{fig:regcoax} was taken to remove the Earth rotation effect (i.e. low frequency effects), and then a histogram taken of the resultant 155,520 measurements (after inverse FFT) to generate the layered histogram plot shown in 
fig.\ref{fig:shnollplotreg}. Layer lines are inserted every 3350 measurements to show a comparison with the Shnoll plot in fig.\ref{fig:shnollplot}. Fig.\ref{fig:shnollplotreg} is remarkably comparable to fig.\ref{fig:shnollplot}, thus suggesting that the Shnoll effect is also present in the coaxial cable  EM anisotropy experiments. The structure observed appears to build up over time instead of cancelling out. It appears slightly noisier but this may be due to the fewer data points obtained than Shnoll (352,980 measurements). The structure observed is found to persist regardless of the time scale used for the phase difference, suggesting that the phenomenon causing this has a fractal nature as depicted in fig.\ref{fig:space6}. If this is indeed caused by a dynamical and fractal 3-space then the persisting structure observed in figures \ref{fig:shnollplot} and \ref{fig:shnollplotreg} correspond to regions of space passing the Earth that have preferred/discrete velocities, and not random ones, as randomly distributed velocities would result in a Poisson distribution, i.e. no features. A likely explanation for this is that the gravitational waves propagating in the 3-space inflow of the Earth or Sun could become phase locked due to the relative locations of massive objects. This would cause reverberation effects, i.e. regions of space which have the same speed and direction, which then repeat over time. The reverberations would be detectable in many experiments such as EM anisotropy, radiation decay, semiconductor noise generation etc. and could in the future be used to further characterise the dynamics of space. 

\begin{figure}[t]
\centering
\includegraphics[scale=0.55]{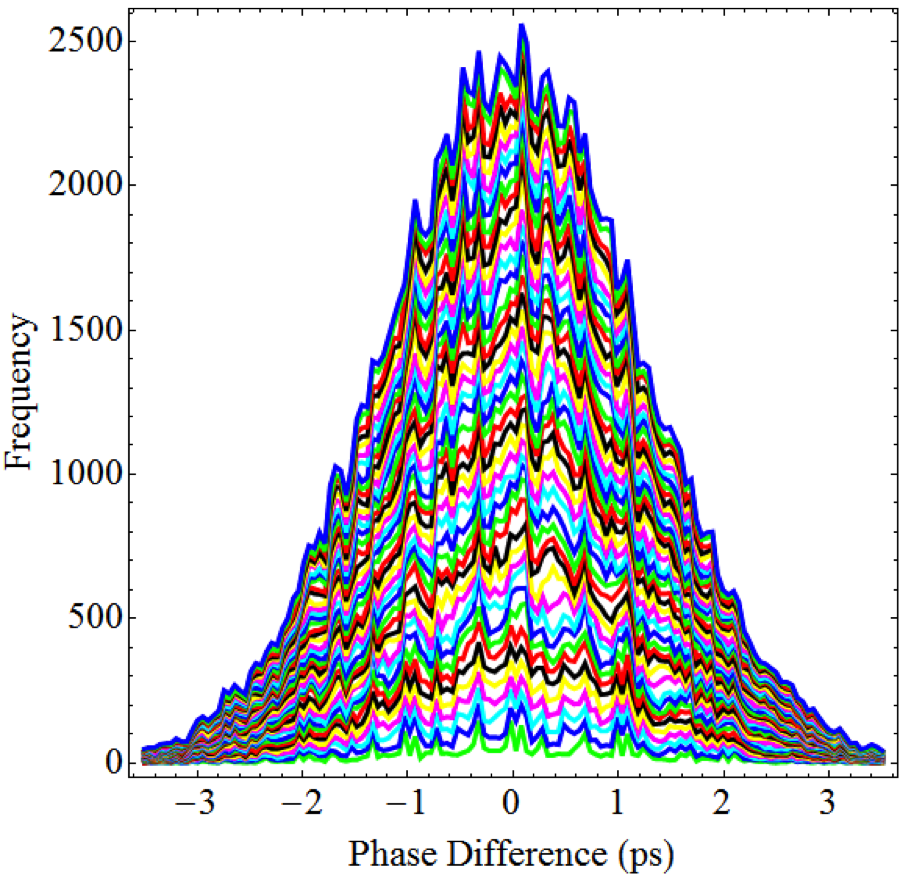}
\vspace{-4mm}\caption{\small{Non-Poisson distribution of 155,520 measurements of the travel time difference (ps) observed between the two coaxial cable circuits of \cite{c1} from Mar 04, 2012 to Mar 12, 2012 in Adelaide. The layered histograms are taken every 3350 measurements to show a comparison with that of fig.\ref{fig:shnollplot}}.
\label{fig:shnollplotreg}}\end{figure}

\begin{figure}[t]
\centering
\includegraphics[scale=0.3]{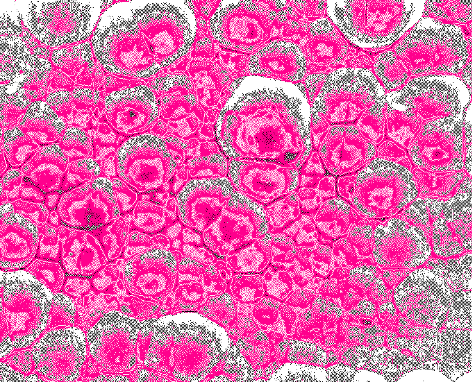}
\vspace{-4mm}\caption{\small{Representation of the fractal wave data as revealing the fractal textured structure of the 3-space, with cells of space having slightly different velocities and continually changing, and moving wrt the earth with a speed of $\sim$500 km/s.}
\label{fig:space6}}\end{figure}

\section{Conclusion}\label{sect:conclusions}

The data from a dual RF coaxial-cable / EM anisotropy - gravitatonal wave  experiment displays the effect Shnoll observed previously in radioactivity experiments. It is suggested that these two experiments (along with other work by Shnoll) are caused by the fractal nature of space, together  with the reverberation effect from gravitational waves,  as predicted  by the Dynamical 3-Space theory.

\section{Acknowledgments}\label{sec:Ack}
Special thanks to Professor Simon Shnoll for permission to use data from his work - see ref \cite{s1} for details.
This report is from the Flinders University Gravitational Wave Detector Project,  funded by an Australian Research Council Discovery Grant: {\it Development and Study of a New Theory of Gravity}.

\end{document}